\documentclass[twocolumn,showpacs,pre,aps]{revtex4}

\usepackage{epsfig}

\def\dt{\Delta t}
\def\dx{\Delta x}
\def\p{\partial}
\def\grad{\nabla}

\def\lk{\lambda_{\bf k}}
\def\d{\Delta}

\begin{document}
\title{Controlling the accuracy of unconditionally 
stable algorithms in Cahn-Hilliard Equation}
\author{Mowei Cheng} 
\author{James A. Warren}
\affiliation{Metallurgy Division and Center for Theoretical and Computational Materials 
Science, National Institute of Standards and Technology, 100 Bureau Drive, Stop 8554,
Gaithersburg, Maryland 20899}


\begin{abstract}
  
  Given an unconditionally stable algorithm for solving the
  Cahn-Hilliard equation, we present a general calculation for an
  analytic time step $\d \tau$ in terms of an algorithmic time step
  $\dt$.  By studying the accumulative multi-step error in Fourier
  space and controlling the error with arbitrary accuracy, we
  determine an improved driving scheme $\dt=At^{2/3}$ and confirm the
  numerical results observed in a previous study \cite{Cheng1}.

\end{abstract}

\maketitle

The Cahn-Hilliard (CH) equation \cite{Cahn1} models the phase separation
that occurs during the quench of a conserved system from a high
temperature isotropic phase into two distinct phases at low
temperatures. The pattern of the two phase regions coarsens as the
time $\tau$ increases, i.e. the length-scale of these regions grows.
At the later stages of this phase ordering process, the dynamics are
dominated by a single length scale, the pattern domain size $L$, which
increases with a power law in time $\tau$, $L(\tau) \sim \tau^{1/3}$
\cite{Bray1}. This power-law growth implies that the motion of the
domain walls becomes extremely slow at late times after a quench since
a typical domain wall speed is $v \sim dL/d\tau \sim \tau^{-2/3}$, and
a typical time scale for the interface to move a distance of order the
interfacial width $\xi$ is of order $\xi/v \sim \tau^{2/3}$.

Since there is no known analytic solution of the Cahn-Hilliard
equation for random initial conditions, computational methods are
necessary for investigation of such systems.  The most straightforward
approach is the Euler algorithm, which must employ a time step
$\dt_{Eu}\sim (\Delta x)^4$ if stability is to be maintained, where
$\dx$ is the lattice spacing.  Additionally, for Cahn-Hilliard
systems, to resolve the interfacial profile, one has to use a lattice
spacing $\dx < \xi$. The Euler fixed time step is suitable for update
near the interface but wastefully accurate in the bulk at late times.
This has been the main challenge of computer simulation of Cahn-Hilliard 
systems.  The recently developed
unconditionally stable algorithm \cite{Eyre1,Vollmayr1,Cheng1}
elegantly overcomes this difficulty. It allows a {\em mode-dependent}
effective time step $\dt_{eff}$ --- a larger effective time step in
the bulk as the domain size gets larger, while keep the effective time
step finite near the interfacial region (see Eq. (\ref{eq:CHdteff_expression})).  
Since the unconditionally stable
algorithm allows no constraints on time step, the main issue is
ensuring the accuracy of the simulation. In a previous study
\cite{Cheng1}, Cheng and Rutenberg numerically demonstrated that the error in
correlations decreases monotonically as $A$ decreases down to $A=0.001$, where
$A$ is a prefactor in the previous driving scheme $\dt=At_s^{2/3}$ and
$t_s$ is the structural time (see below for a precise definition).
However, absence of computational power prevents us from exploring the error
behavior for arbitrarily small $A$. For {\em arbitrary accuracy}, we
need to rigorously prove that the error can be made {\em arbitrarily small}.

To achieve this goal, we must first distinguish two quantities generic
to all numerical algorithms: {\em analytic time step $\d \tau$}
({\em analytic time $\tau$}) and {\em algorithmic time step $\dt$}
({\em algorithmic time $t$}). The former appears in the equation of
motion and represents the time step (time) of the system evolution
governed by the exact solution to the dynamical equations, while the
latter appears in the finite difference scheme and represents the time
step (time) of the system evolution by the computational algorithms.
We now briefly review these two concepts and study why the distinction
has been largely overlooked thus far.

In Euler algorithm, it is not necessary to distinguish the analytic
time step and the algorithmic time step, since there is a threshold on
the time step, and they are always approximately identical (see the
analysis below). On the other hand, semi-implicit algorithms have no
such threshold. An unconditionally stable algorithm, an extension of 
the semi-implicit method, allows for an arbitrarily large time step 
without encountering the numerical instabilities for suitably chosen 
parameters (as determined
using a standard von Neumann stability analysis). Although this
method has been in use for some time, there has been little analytic
study about how to obtain maximal speedup while controlling the
accuracy.  Indeed, all the previous work known to us assumes no difference
between the analytic time step and the algorithmic time step --- one
can only increase the time step modestly and assume the resulting
error are small enough to be ignored.  

In what follows, we perform a general calculation of the analytic
time step $\d \tau$ in terms of the algorithmic time step $\dt$, and
show how this relationship allows one to choose a driving scheme for
arbitrary accuracy.  Concomitantly, we demonstrate that the driving
scheme can be improved to $\dt=At^{2/3}$.  While the calculation
presented is specifically applicable to the Cahn-Hilliard equation,
much of our analysis is general, and should guide subsequent studies
of more complicated systems. For simplicity but without loss of
generality, we restrict our analysis to two dimensions ($2$D).

The Cahn-Hilliard equation can be written as
\begin{equation}
\frac{\p \phi}{\p \tau}=\grad^2 \frac{\delta F}{\delta \phi}
=-\grad^2(\phi+\grad^2\phi-\phi^3),
         \label{eq:CHbasic}
\end{equation}
where the free energy functional is
\begin{equation}
F \equiv \int d^2x \left[|\grad \phi|^2 + \frac{(\phi^2-1)^2}{4}\right],
\end{equation}
and $\phi({\bf x},\tau)$ is a conserved scalar field (such as an
appropriately scaled mass concentration) and the potential has a
double-well structure that the equilibrium values are at $\phi=\pm 1$.
To illustrate and distinguish the analytic time step from the 
algorithmic time step, we first study the exact dynamics of the 
Cahn-Hilliard systems Eq. (\ref{eq:CHbasic}) in Fourier space.  At 
an analytic time $\tau$, the system evolution after an analytic time 
step $\d \tau$ is governed by the Taylor expansion:
\begin{equation}
\phi_k(\tau+\d \tau)=\phi_k(\tau) + \sum_{n=1}^\infty 
\frac{\p^n \phi_k}{\p \tau^n} \frac{\d \tau^n}{n!}.
        \label{eq:CH_Taylor}
\end{equation}
Using the results of the field derivatives in Fourier space Eq.
(\ref{eq:CHfieldscaling}), for a finite $\d \tau$, one finds that all
$n \ge 2$ terms are negligible compared with the $n=1$ term.  So we
obtain the traditional Euler finite difference scheme
\begin{equation}
\phi_k(t+\dt_{Eu})=\phi_k(t)+ \dt_{Eu} \frac{\p {\phi}_k}{\p \tau}, 
        \label{eq:CH_Euler}
\end{equation}
where $\p \phi_k/\p \tau$ is the Fourier transform of $\p \phi/\p
\tau$ in Eq. (\ref{eq:CHbasic}) and is a function of $\phi_k(t)$. We
see that the Euler algorithm uses first order finite differences to
approximate the solution obtained by exact dynamics. On the other
hand, an unconditionally stable algorithm is obtained by an
appropriate semi-implicit
discretization of  Eq. (\ref{eq:CHbasic}) in algorithmic time:
\begin{eqnarray}
\phi_{t+\dt} + (1-a_1) \dt \grad^2 \phi_{t+\dt} + (1-a_2) \dt
\grad^4 \phi_{t+\dt} \nonumber \\
= \phi_t - \dt \grad^2 (a_1 \phi_t + a_2 \grad^2\phi_t - \phi_t^3).
        \label{eq:CHdirect}
\end{eqnarray}
Unconditionally stability is obtained for the choices $a_1>2$ and
$a_2<0.5$ \cite{Vollmayr1}.  Here, $\phi_{t+\dt}$ represents the implicit
terms and $\phi_t$ represents the explicit terms.  We can solve Eq.
(\ref{eq:CHdirect}) directly in Fourier space and obtain
\begin{equation}
\phi_k(t+\dt)= \phi_k(t) + \dt_{eff}(k,\dt) \frac{\p {\phi}_k}{\p \tau},
        \label{eq:CHdteff}
\end{equation}
where the $k$-dependent effective time step is
\begin{equation}
        \dt_{eff}(k,\dt) \equiv \frac{\dt}{1-\dt \lk [(a_1-1) + (a_2-1) \lk]},
        \label{eq:CHdteff_expression}
\end{equation}
and $\lk=-k^2$ is the Fourier-transformed Laplacian.  The Euler
algorithm has a mode-independent fixed time step to update the system
in Fourier space, but, as Eq. (\ref{eq:CHdteff_expression}) reveals,
the unconditionally stable algorithm has a mode-dependent effective
time step $\dt_{eff}(k,\dt)$. A direct comparison of Eq.
(\ref{eq:CH_Taylor}) and Eq.  (\ref{eq:CH_Euler}) yields that the 
analytic time step $\d \tau$ is always a good approximation of
the algorithmic time step $\dt_{Eu}$ in Euler algorithm.
However, for the unconditionally stable algorithm, a comparison of Eq.
(\ref{eq:CH_Taylor}) and Eq.  (\ref{eq:CHdteff}) does not give a
straightforward relation between $\d \tau$ and $\dt$, i.e., we do not
know what $\d \tau$ corresponds to $\dt$.  In what follows we explore
the relationship between these two time steps in CH equation, and the
consequences this relationship has on the accuracy of the solution
method. The steps of our procedure shown in {\em italics}.

{\em Calculate the analytic time and time step}: We now calculate
the analytic time step $\d \tau$ in terms of an algorithmic time
step $\dt$. Cahn-Hilliard systems are purely dissipative systems ---
the energy density $E$ monotonically decreases with the analytic
time with the relation $E \propto \tau^{-1/3}$ \cite{Bray1}.  Without
such a relationship between a physical quantity and the analytic time,
the analysis  that is performed below cannot proceed, and thus
progress in applying these methods to other models hinges on the
physical insights needed to obtain such relationships (in this case
the so-called ``scaling hypothesis''). The analytic time is conveniently
calculated in terms of the monotonically decaying energy density $E$:
$\tau = B/E^3$, where the prefactor $B$ can be numerically determined
by requiring $\d \tau=\dt$ as $\dt \to 0$ in the late-time scaling
regime since our unconditionally stable algorithm is arbitrarily
accurate as $\dt \to 0$. Note that the calculation here is identical
to the calculation of the {\em structural time} $t_s$ in a previous
study \cite{Cheng1} since the structural time is just another 
representation of the {\em analytic time}.

We can calculate the analytic time step by differentiating $\tau$
with respect to $E$:
\begin{equation}
\d \tau=-3B \frac{\Delta E}{E^4}=-3 \Delta E \frac{\tau^{4/3}}{B^{1/3}}, 
\end{equation}
and $\Delta E$ can be calculated by integrating $\Delta E$ from each 
Fourier mode:
\begin{eqnarray}
\Delta E
&\approx& \int_0^{1/\xi} d^2k \frac{1}{(2\pi)^2}
\left\langle \left(\frac{\delta F}{\delta \phi_k}\right) \Delta \phi_k 
\right\rangle
\nonumber \\
&=&-\int_0^{1/\xi} d^2k \frac{1}{(2\pi k)^2} \dt_{eff}(k,\dt) T_k,
        \label{eq:CHde}
\end{eqnarray}
where the time derivative $\p \phi_{-k}/\p \tau=-k^2 \delta F/\delta \phi_k$
from Eq. (\ref{eq:CHbasic}) and 
$\Delta \phi_k=\phi_k(t+\dt)-\phi_k(t)=\dt_{eff} \p \phi_k/\p \tau$ from 
Eq. (\ref{eq:CHdteff}) are used, and $T_k$ is the time-derivative correlation 
function \cite{Bray1,Bray2,Rutenberg1}, and has a natural scaling form given by
\begin{eqnarray}
T_k \equiv \left \langle \frac{\p \phi_k}{\p \tau} 
\frac{\p \phi_{-k}}{\p \tau} \right \rangle
= \left( \frac{dL}{d \tau} \right)^2 h(kL) 
= \frac{L_0^2h(kL)}{9 \tau^{4/3}},
\end{eqnarray}
where $L=L_0\tau^{1/3}$,  $h(x) = C/x$   is the $2$D scaling function
\cite{Rutenberg1} as $x \gg 1$, and $L_0$ and $C$ are constants. We 
can then solve for $\Delta E$ in Eq. (\ref{eq:CHde}) and for the 
analytic time step $\d \tau$:
\begin{eqnarray}
\d \tau
&=& \frac{L_0^2 \dt}{6\pi B^{1/3}} \int_0^\infty \frac{dx}{x} 
\frac{h(x)}{1+\dt(a_1-1)x^2/L^2} \nonumber \\
&=& \frac{CL_0^2 \dt}{6\pi B^{1/3}} \int_0^\infty \frac{dx}{x^2(1+D \ell^2 x^2)},
\end{eqnarray}
where $x=kL$, $D=(a_1-1)/L_0^2$ and $\ell=\sqrt{\dt/\tau^{2/3}}$. Solving the
integral, we obtain that,
\begin{eqnarray}
\d \tau=\dt \left[1 - \zeta \ell + \mathcal{O}(\ell^2) \right],
        \label{eq:CHprediction-dts-dt}
\end{eqnarray}
where $\zeta=L_0 C \sqrt{a_1-1}/(12 B^{1/3})$. The above formula is
the central result of this manuscript, and implies that $\d \tau \le
\dt$ in general.  We now explore how to use this result to obtain
an accelerated algorithm.

{\em Scaling of field derivatives in Fourier space}:
In order to explore the accuracy of accelerated algorithms in Fourier space,
it is necessary to know the scaling of field derivatives both in the bulk 
(where $k \sim 1/L$) and near the interface (where $k \sim 1/\xi$).
The structure factor $S(k) = \langle|\phi_k|^2\rangle =L^2g(kL)$, where
$g(kL) \sim 1$ as $k \sim 1/L$ and $g(kL) \sim (kL)^{-3} \sim L^{-3}$ as $k \sim
1/\xi$ \cite{Bray1}. Therefore we obtain
\begin{eqnarray}
\phi_k \sim \cases{\tau^{1/3} & \quad as $k
\sim 1/L$ \cr \tau^{-1/6} & \quad as $k \sim 1/\xi$ \cr}
\end{eqnarray}

Previous studies \cite{Bray1,Rutenberg1} showed that
$\p \phi_k/\p \tau = (dL/d \tau) k \phi_k$ as $kL \gg 1$, so we obtain the form for the
time-derivative correlation function 
$T(k)=\langle|\p \phi_k/\p t|^2\rangle=(dL/d \tau)^2k^2\langle|\phi_k|^2\rangle=(dL/d \tau)^2h_1(kL)$,
where the scaling function $h_1(kL)=k^2L^2g(kL) \sim 1$ as $k \sim 1/L$, and
$h_1(kL) \sim (kL)^{-1} \sim L^{-1}$ as $k \sim 1/\xi$. Therefore we obtain
\begin{eqnarray}
\frac{\p \phi_k}{\p \tau} \sim
\cases{\tau^{-2/3} & \quad as $k \sim 1/L$ \cr \tau^{-5/6} & \quad as $k \sim
1/\xi$
\cr}
\end{eqnarray}

The generalization of higher order time-derivative correlations is
$\langle|\p^n \phi_k/\p \tau^n |^2\rangle 
\sim (dL/d \tau)^2k^2\langle|\p^{n-1} \phi_k/\p t^{n-1} |^2\rangle$,
where ``$\sim$'' indicates that generally the left hand side may not exactly be 
equal to the right hand side. Applying this relation will yield
$\langle|\p^n \phi_k/\p \tau^n|^2\rangle \sim (dL/d \tau)^{2n}L^{2-2n}h_n(kL)$,
where $h_n(kL)=k^2L^2 h_{n-1}(kL) \sim (kL)^{2n-3} \sim 1$ as $k \sim 1/L$, and
$h_n(kL) \sim (kL)^{2n-3} \sim L^{2n-3}$ as $k \sim 1/\xi$. Therefore we have
\begin{eqnarray}
\frac{\p^n \phi_k}{\p \tau^n} \sim
\cases{\tau^{-n+1/3} & \quad as $k \sim 1/L$ \cr \tau^{-2n/3-1/6} & \quad as $k
\sim
1/\xi$ \cr}
        \label{eq:CHfieldscaling}
\end{eqnarray}
The above expression is valid for $n \ge 0$ for conserved two dimensional scalar 
order parameter(s).

{\em Determine the driving scheme for arbitrary accuracy}: Next, we
determine the driving scheme for arbitrary accuracy in terms of the
Fourier space error. Before we study the error, we must first distinguish
the error in the bulk and the error near the interface. Eq.
(\ref{eq:CHdteff_expression}) implies $\dt_{eff} \sim \tau^{2/3}$ as
$k \sim 1/L$ and $\dt_{eff} \sim const.$ as $k \sim 1/\xi$, we obtain
that the ratio of the single step field update with respect to the field 
$\Delta \phi_k/\phi_k \sim (\dt_{eff} \p \phi_k/\p \tau)/\phi_k$ is of order
$\mathcal{O}(\tau^{-1/3})$ as $k \sim 1/L$ and
$\mathcal{O}(\tau^{-2/3})$ as $k \sim 1/\xi$.  Therefore the error near
the interface is negligible compared with the error in the bulk, and
we will only study the error of those modes where $k \sim 1/L$.

In Fourier space, we compare the field evolved by an unconditionally
stable algorithm to the exact dynamics evolved by the {\em same amount
  of energy}.  Using this criterion we obtain the Fourier space single
step error
\begin{eqnarray}
\Delta\phi_k^s 
&\equiv& \phi_k(t+\dt)-\phi_k(t+\d \tau) 
\nonumber \\
&=& \left( \dt_{eff} - \d \tau \right) \frac{\p \phi_k}{\p \tau}
-\sum_{n=2}^\infty \frac{\p^n \phi_k}{\p \tau^n} \frac{\d \tau^n}{n!} 
\nonumber \\
&\sim& \frac{1}{1+D \ell^2} \left[\zeta \ell^3 + \mathcal{O}(\ell^4) \right]
        \label{eq:single_step_error}
\end{eqnarray}
where Eq. (\ref{eq:CHprediction-dts-dt}) and $\p \phi_k/\p \tau \sim
\tau^{-2/3}$ as $k \sim 1/L$ are used. The values of $\zeta$ and $D$
are finite. Assuming the algorithmic time step $\dt=A \tau^{\beta}$,
then $\ell=\sqrt{A}\tau^{\beta/2-1/3}$.  In order to obtain arbitrary
accuracy for $\Delta\phi_k^s$ at arbitrarily large $\tau$, we require
that $\beta=2/3$ since $\beta>2/3$ will make the error uncontrolled
(arbitrarily large) at arbitrarily large $\tau$, and $\beta<2/3$ will
make the algorithm wastefully accurate (error is always zero) at
arbitrarily large $\tau$.  $A$ is then selected so that a desired
accuracy is obtained. Thus, $\dt=A \tau^{2/3}$ and $\Delta\phi_k^s
\sim \mathcal{O}(\ell^3) \sim \mathcal{O}(A^{3/2})$.

For small $A$, Eq. (\ref{eq:CHprediction-dts-dt}) implies
that $\tau \approx t \left(1 - \zeta \sqrt{A} \right)$. Therefore we 
can express the algorithmic time step $\dt$ in terms of algorithmic 
time $t$:
\begin{eqnarray}
\dt=A\left(1 - \zeta \sqrt{A} \right)^{2/3}  t^{2/3} \approx At^{2/3}.
\end{eqnarray}
Writing the driving algorithmic time step in terms of algorithmic time $t$ 
instead of the analytic time $\tau$ has the computational advantage
of avoiding an intermediate calculation of  $\tau$ at each update,
and thus makes the computational implementation more straightforward.

{\em Accuracy in correlations}: Lastly, we analytically confirm the
numerical results in a previous study \cite{Cheng1} that the error in
structure factor scales as $\sqrt{A}$. The Fourier space single-step
error Eq. (\ref{eq:single_step_error}) will at worst accumulate with
each update.  For a small $A$, evolving to $\tau$ with time-step $\dt
= At^{2/3} \approx A \tau^{2/3} \approx \d \tau \sim d\tau/dn$
requires a number of steps
\begin{eqnarray}
n=\int dn\sim\int_0^{\tau} \frac{d\tau}{A\tau^{2/3}} = \frac{3\tau^{1/3}}{A}.
\end{eqnarray}
Therefore, at $\tau$, we obtain the upper bound on the Fourier space multi-step error:
\begin{eqnarray}
\Delta\phi_k^m \sim \Delta\phi_k^s n 
\sim \frac{3A^{3/2} \tau^{1/3}}{A} \sim L \sqrt{A}.
\end{eqnarray}
We can use this to bound the error in the scaled structure factor
$g(kL)=\langle|\phi_k|^2\rangle/L^2$. As was investigated in our
previous numerical studies \cite{Cheng1}, this quantity is simply the magnitude
of the difference between the structure factor obtained using an
unconditionally stable algorithm with one using the exact dynamics at
the {\em same energy}. As $k \sim 1/L$ (in the bulk), we obtain the
maximum error:
\begin{eqnarray}
\Delta g_{max} \approx \frac{2\Delta\phi_k^m \phi_k}{L^2} \sim \sqrt{A},
        \label{eq:structureg}
\end{eqnarray}
where $\phi_k \sim L$ as $k \sim 1/L$ is used. 
Eq. (\ref{eq:structureg}) is precisely the same as the results obtained in
our previous, solely numerical, study \cite{Cheng1}.  Thus, the error
produced in the bulk dominates the total error, as it decays much
slower than the error produced near the interface.
This error accumulates over time and results the error in the
structure factor scaling as $\sqrt{A}$.

In summary, we have analyzed numerical methods for solving the
Cahn-Hilliard Equation.  By explicitly distinguishing the analytic and
algorithmic time steps, we have developed a relation between them, and
have obtained an optimal driving scheme $\dt=At^{2/3}$ under the
requirement of arbitrary accuracy.  With this driving scheme, we have
proved that the upper bound of the multi-step error in structure
factor scales as $\sqrt{A}$, a result obtained by numerical methods in
a previous study \cite{Cheng1}.  We note that the argument developed
herein is founded, ultimately, on the physical relationship between
the domain size and the analytic time (based itself on the scaling
hypothesis).  For systems where such relationships exist, or can be
derived, we expect that this analysis should generalize to other
systems, such as newly developed Phase Field Crystal model
\cite{Elder1,Elder2}. We hope to report this work in a subsequent
paper.

MC would like to acknowledge Andrew Rutenberg for valuable discussions on 
this work.




\begin{thebibliography}{}

\bibitem{Cahn1} J. W. Cahn and J. E. Hilliard, J. Chem. Phys. {\bf 28}, 258 (1958).
\bibitem{Bray1} A. J. Bray, Adv. Phys. {\bf 43}, 357 (1994).
\bibitem{Eyre1} D. J. Eyre, in {\em Computational and Mathematical Models of 
Microstructural Evolution}, edited by J. W. Bullard {\em et al.} (The Material 
Research Society, Warrendale, PA, 1998), pp. 39-46.
\bibitem{Vollmayr1} B. P. Vollmayr-Lee and A. D. Rutenberg, Phys. Rev. E {\bf 68}, 66703 (2003).
\bibitem{Cheng1} M. Cheng and A. D. Rutenberg, Phys. Rev. E {\bf 72}, 055701(R) (2005).
\bibitem{Bray2} A. J. Bray and A. D. Rutenberg, Phys. Rev. E {\bf 49}, R27 (1994).
\bibitem{Rutenberg1} A. D. Rutenberg and A. J. Bray, Phys. Rev. E {\bf 51}, 5499 (1995).
\bibitem{Elder1}K. R. Elder, M. Katakowski, M. Haataja, and M. Grant, Phys. Rev. Lett. {\bf 88},
245701 (2002).
\bibitem{Elder2}K. R. Elder and M. Grant, Phys. Rev. E {\bf 70}, 051605 (2004).

\end{thebibliography}
\end{document}